# Tellegen responses in metamaterials

Qingdong Yang [1†], Xinhua Wen [1†], Zhongfu Li[1†], Oubo You[1], Shuang Zhang[1,2,3]∗

**Affiliations:**

[1] New Cornerstone Science Laboratory, Department of Physics, University of Hong Kong; 999077, Hong Kong, China

[2] Department of Electrical & Electronic Engineering, University of Hong Kong; 999077, Hong Kong, China.

[3] Materials Innovation Institute for Life Sciences and Energy (MILES), HKU-SIRI, Shenzhen, P.R. China

*Corresponding author. Email: shuzhang@hku.hk

†These authors contributed equally to this work.

# Abstract

Tellegen medium has long been a topic of debate, with its existence being contested over several decades. It was first proposed by Tellegen in 1948 and is characterized by a real-valued cross coupling between electric and magnetic responses, distinguishing it from the well-known chiral medium that has imaginary coupling coefficients. Significantly, Tellegen responses are closely linked to axion dynamics, an extensively studied subject in condensed matter physics. Here, we report the realization of Tellegen metamaterials in the microwave region through a judicious combination of subwavelength metallic resonators, gyromagnetic materials, and permanent magnet discs. We observe the key signature of the Tellegen response – a Kerr rotation for reflected wave, while the polarization remains the same in the transmission direction. The retrieved effective Tellegen parameter is several orders of magnitude greater than that of natural materials. Our work opens door to a variety of nonreciprocal photonic devices and may provide a platform for studying axion physics.

# Introduction

Tellegen media, characterized by a real-valued cross-coupling between electric and magnetic responses, are first theoretically proposed by Tellegen in 1948 for designing an electromagnetic gyrator[1]. Ever since then, Tellegen media has drawn significant attention and sparked extensive discussions regarding its existence in electromagnetics[2-7]. The Tellegen coupling, as a subclass of the bianisotropic coupling, renders additional degrees of freedom for manipulating the light-matter interactions and controlling the wave polarizationc[8,9]. When an electromagnetic wave interacts with a Tellegen material, angular momentum is transferred to reflected waves, inducing a rotation of the polarization plane of the reflected wave, referred to as the Kerr effect[10]. Differing from extensively studied reciprocal chiral media[11-13], Tellegen media possess a nonreciprocal nature due to the broken parity symmetry and time-reversal symmetry. Therefore, Tellegen materials offer the promise of nonreciprocal wave propagation, with practical applications such as electromagnetic isolators and nonreciprocal twist polarizers[14,15]. In addition, the Tellegen response is closely associated with axion electrodynamics, which has drawn great attention due to its connections with topological insulators, high-energy physics, and dark matters[16-19]. The electrodynamic equations of a media with a pure Tellegen response have the same form as those of the axion media[20-22], making it an exciting opportunity to explore axion-related physics with Tellegen materials. Despite the substantial interest and significance of the Tellegen response in both applied and fundamental physics, most of the research about the Tellegen response has remained confined to theoretical exploration over the past two decades[23-29]. A ferrite-based structure has been proposed as a Tellegen particle in the microwave regime, based on the measurement of the amplitude of the cross-polarized wave in a waveguide[29]. However, it was later argued that such a measurement in the waveguide was not convincing, as no definite and stable polarizability parameters characterizing the magnetoelectric properties could be found[24]. To date, a decisive demonstration of Tellegen media exhibiting their key features still remains elusive.

In this work, we report the experimental realization of a metamaterial with a pure Tellegen response, and the first observation of a gigantic Kerr rotation that is several orders of magnitude stronger than that observed in natural materials. Specifically, combining gyromagnetic materials and split-ring resonators, a Tellegen metamaterial with in-phase magnetoelectric coupling is designed and fabricated. When an electromagnetic wave impinges on the Tellegen metamaterial slab, the polarization of the reflection wave is rotated (i.e., the Kerr effect), while the transmission wave remains the same polarization as the incident field, serving as the most key characteristics of a pure Tellegen response that has been missing so far. Notably, giant Kerr rotation reaching 90° is observed, indicating a strong effective Tellegen coupling parameter several orders of magnitude greater than that of intrinsic magnetoelectric materials (e.g., $Cr_2O_3$[23]). It is noted that the pure Tellegen response is closely associated with axion electrodynamics. Furthermore, the measured magnetoelectric nearfield distribution within a metamaterial unit-cell provides more evidence supporting the rotation of polarization response in the far-field. The experimental realization of Tellegen metamaterials not only holds promise for various practical applications, such as electromagnetic isolators and nonreciprocal twist polarizers, but also opens up significant opportunities for the experimental exploration of novel concepts related to axion electrodynamics.

## Results

**Tellegen media's polarizability properties**

We start with a general description of the polarizability of bianisotropic meta-atoms, which plays a pivotal role in governing the electromagnetic response of bianisotropic metamaterials. For a general bianisotropic meta-atom, the induced electric dipole moments **p** and magnetic dipole moments **m** can be related to the incident electromagnetic fields $\mathbf{E}_{in}$ and $\mathbf{H}_{in}$ by a polarizability tensor[3,30]:

$$\begin{pmatrix} \mathbf{p} \\ \mathbf{m} \end{pmatrix} = \begin{pmatrix} \bar{\bar{\alpha}}_{ee} & \bar{\bar{\alpha}}_{em} \\ \bar{\bar{\alpha}}_{me} & \bar{\bar{\alpha}}_{mm} \end{pmatrix} \begin{pmatrix} \mathbf{E}_{in} \\ \mathbf{H}_{in} \end{pmatrix} \qquad (1)$$

where $\bar{\bar{\alpha}}_{ee}$ and $\bar{\bar{\alpha}}_{mm}$ denote the electric and magnetic polarizability tensors respectively; $\bar{\bar{\alpha}}_{em}$ and $\bar{\bar{\alpha}}_{me}$ respectively represent the electromagnetic and magnetoelectric polarizability tensors, describing the cross-coupling between the electric and magnetic responses. The isotropic bianisotropic coupling can be categorized into four distinct classes – two that are reciprocal (chiral and omega), and two that are nonreciprocal (Tellegen and moving). Each category exhibits unique transmission and reflection in response to incident waves, determined by the phase and polarization of induced dipoles[31-33]. Tellegen materials represents a specific class of bianisotropic coupling that is characterized by a real-valued (in-phase) cross-coupling polarizability.

To gain a comprehensive understanding of Tellegen response, we provide a comparative illustration depicting the electromagnetic response between a Tellegen metamaterial slab and an extensively studied chiral metamaterial slab in Fig. 1[34-36]. We assume that the metamaterial slabs are positioned in the $x$-$y$ plane, and they are illuminated by a normally incident $y$-polarized plane wave. For a chiral meta-atom characterized by an imaginary-valued cross-coupling polarizability, the electric field $E_y$ (magnetic field $H_x$) induces a magnetic dipole $-im_y$ (an electric dipole $ip_x$), as illustrated in the left inset of Fig. 1a. Notably, the presence of the imaginary unit $i$ indicates that the induced magnetic and electric dipole moments exhibit an additional $\pi/2$ phase shift with respect to incident fields. In the chiral composite slab, the two dipole moments both emit scattered waves of orthogonal polarization to that of incident wave, which constructively interference in the forward direction, giving rise to a rotation in the polarization of the transmitted wave, commonly referred to as the Faraday effect. Consequently, as depicted in Fig. 1a, the incident wave, polarized along the $y$-axis, transforms into a linear polarized transmitted wave, while the polarization plane of the reflected wave remains unchanged due to the destructive interference of the $x$-polarized scattered wave.

In contrast, in the case of a Tellegen meta-atom, the induced magnetic (electric) dipole moments align parallel to the incident electric (magnetic) field (the right inset of Fig. 1b) with an in-phase coupling. In the metamaterial slab with pure Tellegen response, the effective electromagnetic response can be expressed as the following constitutive equations[37]:

$$\begin{pmatrix} D_x \\ D_y \end{pmatrix} = \varepsilon_0 \varepsilon \begin{pmatrix} E_x \\ E_y \end{pmatrix} + \frac{1}{c_0} \begin{pmatrix} \xi & 0 \\ 0 & \xi \end{pmatrix} \begin{pmatrix} H_x \\ H_y \end{pmatrix} \qquad (2)$$

$$\begin{pmatrix} B_x \\ B_y \end{pmatrix} = \mu_0 \mu \begin{pmatrix} H_x \\ H_y \end{pmatrix} + \frac{1}{c_0} \begin{pmatrix} \xi & 0 \\ 0 & \xi \end{pmatrix} \begin{pmatrix} E_x \\ E_y \end{pmatrix}$$

where $\varepsilon$ and $\mu$ are the effective permittivity and permeability, and $\xi$ represents the effective Tellegen parameter. The Tellegen response, featuring a real-valued cross-coupling term, violates the Onsager-Casimir symmetry relations while holding the Hermitian requirements that the induced magnetic dipole $m_y$ and electric dipole $p_x$ have the same phase as the incident fields. In contrast to the chiral slabs, scattered waves of orthogonal polarization interfere constructively (destructively) in the backward (forward) direction, resulting in a rotation of the polarization plane of the reflected wave. This effect is called the Kerr effect. It should be noted that the metamaterial slab with pure Tellegen response exhibits the Kerr effect for arbitrary polarization of linearly polarized incident waves and ensures that the transmitted wave passing through the Tellegen metamaterial slab retains the same polarization as the incident wave as schematically illustrated in Fig. 1b.

Interestingly, the constitutive relations of the Tellegen media in Eq. (2) can be rewritten as:

$$\boldsymbol{H} = \frac{\boldsymbol{B}}{\mu_0 \mu} - \alpha \vartheta \boldsymbol{E}, \qquad (3)$$

$$\boldsymbol{D} = \varepsilon_0 \varepsilon' \boldsymbol{E} + \alpha \vartheta \boldsymbol{B}.$$

where the field vectors $\boldsymbol{B}, \boldsymbol{H}, \boldsymbol{D}, \boldsymbol{E}$ contain both $x$ and $y$ components. $\alpha$ is coupling constant. Eq. (3) presents the constitutive relations of the axion media [20,38]. The axion-coupling term $\vartheta$ is related to the Tellegen coupling term by $\alpha \vartheta = \xi / c_0 \mu_0 \mu$. The effective permittivity of the axion media is modified as $\varepsilon_0 \varepsilon' = \varepsilon_0 \varepsilon - \xi^2 / c_0^2 \mu_0 \mu$, while the effective permeability remains the same as that of the Tellegen media. The presence of the axion-coupling term modifies the Maxwell equations within axion media, giving an additional term in Lagrangian density (Supplementary Materials Sec. 1). Consequently, the Tellegen metamaterial, which embodies the equations of axion electrodynamics, serves as a practical platform for investigating the intriguing associated phenomena. In particular, at the interface between the air and Tellegen metamaterial slab, the sudden change of the axion field affects the electromagnetic response, giving rise to the distinctive Kerr effect, which is the fingerprint of the axion response[18].

**Tellegen metamaterial design**

To achieve pure Tellegen coupling as described in Eq. (2), we propose a Tellegen metamaterial design that incorporates saddle-shaped connective metallic coils and rods of gyromagnetic materials biased by external magnetic fields generated by locally positioned magnets, as schematically shown in Fig. 2a. Specifically, each sub-unit consists of an Yttrium-iron-garnet (YIG) rod (black cylinder) placed in the center of the saddle-shaped connective metallic coil (yellow color), biased by a small positive magnet (bright disc). Each unit-cell comprises four sub-units, with two of them biased by opposite external magnetic fields relative to the other two, as indicated by the blue arrows in Fig. 2a. The saddle-shaped connective metallic coils, with four-fold rotational symmetry, can be effectively regarded as two orthogonal splitting resonator rings (SRRs), which are represented by red and blue metal strips, respectively, as shown in Fig. 2b. Similarly, the inverted sub-particle corresponds to flipped SRRs, as shown in the right panel of Fig. 2b.

The schematic in Fig. 2c elucidates the underlying mechanism responsible for generating the pure Tellegen response within the unit-cell in Fig. 2a. When a plane wave with $y$-polarization is incident onto the sub-unit structure, the $x$-oriented magnetic field $H_x$ induces an oscillating magnetic dipole moment $-im_y$ (upper panel), due to the gyromagnetic effect (i.e., the antisymmetric part of the YIG permeability tensor). According to the Lenz-Faraday law, the magnetic dipole moment excites the electric current in the SRR and causes opposite charge accumulation in the two vertical arms, leading to an *x*-directed electric dipole $p_x$ (lower panel). The induced electric dipole is collinear with the incident magnetic field, representing an in-phase cross-coupling, which is the essence of the Tellegen response. As depicted in Fig. 2c, the sub-units, biased by opposite magnetic fields, induce magnetic dipole moments of opposite polarities (denoted by $-im_y$ and $im_y$ respectively), while giving rise to the same electric dipoles $p_x$. The *y*-oriented electrical field $E_y$ induce $m_y$ in a similar manner, as shown in Supplementary Fig. 1. Consequently, the metamaterial unit-cell, consisting of two sets of sub-units with reversed orientation, effectively cancels out unwanted responses such as the gyromagnetic and omega response, and provides pure and isotropic Tellegen response in the *x-y* plane.

**Experiment measurement**

A Tellegen metamaterial slab is constructed by arranging the designed metamaterial unit-cell structure periodically along the *x-* and *y*-directions with a period of L=17.2 mm. The sample was fabricated using the Printed Circuit Board (PCB) technology. The saddle-shaped connective metallic coil is embedded inside Teflon with a relative permittivity of 2.72 and loss tangent of 0.008. The YIG rods and permanent magnets are inserted into the circular apertures drilled into the fabricated circuit board, as shown in the photograph presented in the upper panel of Fig. 3a. The red frame marks the actual unit-cell structure, wherein the permanent magnets of two sub-particles have reversed orientation to provide opposite magnetic bias. For the experimental measurement of the transmission, the sample is positioned between two linearly polarized wideband horn antennas connected to the ports of a Vector Network Analyzer (VNA), as shown in Fig. 3a. The incident wave from the source horn antenna is polarized in the *y*-direction and normally impinged on the metamaterial slab. By aligning the receiving antenna parallel (perpendicular) to the source antenna, we can measure the co-polarized (cross-polarized) transmission signals. Similarly, by placing the probe antenna on the same side as the source antenna, we can measure both co- and cross-polarized reflected signals. More details about the sample are provided in the Supplementary Materials Sec. 2.

The measured reflection and transmission intensity spectra for both the co- and cross-polarization are shown in Fig. 3b. The reflection spectra reveal that the reflection wave undergoes polarization rotation across a broad frequency range, indicating the presence of the Kerr effect induced by the Tellegen response. At the resonant frequency of 9.5 GHz, the cross-polarized reflectivity reaches its maximum, while the co-polarized reflectivity is almost zero, representing a near-complete cross polarization conversion in reflection. This indicates a strong Tellegen response at the resonance frequency. Furthermore, as shown by Fig. 3b, the transmission wave remains the same polarization as the incident wave, as evidenced by the near-zero cross-polarization transmissivity observed across the entire frequency range of interest, which is another key signature of a pure Tellegen response. We also perform simulation on the reflection and transmission, with the results shown in Fig. 3c. The simulation is conducted using the commercial software COMSOL, employing a periodic unit-cell structure. Despite some discrepancies in the higher frequency range due to the finite size of the sample, the simulation results show good agreement with the measured spectra, further validating the presence of pure Tellegen response of the metamaterial slab.

Both the simulated and measured polarization rotation of reflected light are provided in Fig. 3d, which confirm the rotation of the reflected light across a wide range of frequencies. The simulated and measured polarization states of the reflected beam at the specific frequencies are shown in. Fig. 3f-g. Due to the presence of loss in the real structure, the polarization state of the reflected beam slightly deviates from the linear polarization and become elliptically polarized. Nonetheless, one can still use the major axis to characterize the polarization rotation, as explained in detail in Supplementary Materials Sec. 3. Near the reflectivity peak at 9.5 GHz, the Kerr rotation is approximately 90° for both simulation and measurement, confirming a near-complete cross polarization conversion from *y*-polarized incident wave to *x*-polarized reflected wave. The rotation angle reaches maximum at a slightly higher frequency than the resonance for both the simulation and experiment. The measured rotation angle is several orders of magnitude greater than that of intrinsic magnetoelectric materials such as $Cr_2O_3$.

The polarization rotation angle of reflected light is closely related to the Tellegen coupling parameters [39]:

$$\tan\theta_{\text{Kerr}} = \frac{2\xi}{\varepsilon - \mu} \qquad (4)$$

Notably, the induced Kerr rotation $\theta_{Kerr}$ is independent of the thickness of the metamaterial slab. The effective Tellegen parameter are retrieved from the transmission and reflectance, as shown in Fig. 3e. The retrieved parameter from both the experimental and simulation results reach its peak at the resonance frequency. Around the resonance frequency, the measured Tellegen parameter $\xi$ is approximately 3.4, several orders of magnitude greater than that of $Cr_2O_3$. Such a substantial Tellegen coupling can be attributed to the engineered resonance of the metamaterial which significantly increases the interaction between the SRR and the YIG rod. Subsequently, we obtain a very large effective axion-coupling term $\vartheta$ at the resonance frequency (Supplementary Fig. 3). Hence the demonstrated Tellegen metamaterial provides a promising platform for exploring axion-related physics.

**Near-field characterization**

We further study the magnetoelectric response of the Tellegen metamaterial via near-field characterization at its surface on the transmission side. Figure 4 shows the $E_z$ component of the nearfield distributions on an *x-y* plane at the transmission side of the sample where the circular metal coils are indicated by the dashed circles. At the frequency of 8.7 GHz, which is away from the resonance frequency, the Tellegen coupling is negligible (Fig. 3d), and the incident field $E_y$ predominantly excites the electric current in the SRRs oriented in the *y*-direction, inducing charge accumulation and consequently electric dipoles oscillating in the *y*-direction. At the resonance frequency of 9.5 GHz, the strong Tellegen coupling of the metamaterial induces the electric dipole oscillating in the *x*-direction. As a result, the superposition of the electric dipoles oscillating in the *y*-direction and *x*-direction leads to an overall nearfield oscillating at a tilted angle with respect to the *y*-axis (Fig. 4b). At a higher frequency of 10.2 GHz, beyond the resonance frequency, the Tellegen coupling starts to vanish, leading to a nearfield $E_z$ predominantly oscillating in the *y* direction (Fig. 4c). The measured near-field distribution is shown in the lower panels of Fig. 4, showing good agreement with the simulated results. Therefore, the microscopic probing of the magnetoelectric nearfield provides further evidence of the Tellegen response. Interestingly, despite that the electric dipoles oscillating at a tilted angle close to the resonance frequency, the transmission should preserve the same polarization as the incident wave, which is protected by the combination of the inversion symmetry and Onsager–Casimir relations (Supplementary Materials Sec. 4). However, lacking any symmetry protection, the polarization of reflected wave is changed as a result of nonreciprocal interaction of incident wave with the metamaterial.

## Conclusion

In summary, we have proposed and experimentally demonstrated a metamaterial slab that exhibits a pure Tellegen response. Such a Tellegen metamaterial is constructed by assembling resonant metallic coils, YIG rods and permanent magnets in a judicious manner, such that the Tellegen response is dramatically enhanced and meanwhile other bianisotropic effects are cancelled. By measuring both the co- and cross-polarized reflection and transmission, we observe a giant polarization rotation in reflection, with a Kerr rotation reaching 90° at the resonance frequency. Meanwhile, the transmission wave remains the same polarization as the incident wave, characteristic of a pure Tellegen response. The extracted effective Tellegen coupling term is several orders of magnitude larger than that of natural materials. Our study bridges the gap in the experimental realization of Tellegen metamaterial, which has significant implications for the understanding of nonreciprocal magnetoelectric response and holds promise for various applications like electromagnetic isolators and nonreciprocal twist polarizers. Furthermore, the Tellegen metamaterial assumes particular importance in the realm of fundamental physics, unlocking the potential for experimental exploration of intriguing and exotic concepts of axion electrodynamics.

## Methods

**Simulation:** We simulated the transmission and reflection spectra of the device using the commercial software COMSOL Multiphysics based on the finite element method (FEM). The relative permeability tensor of the gyromagnetic materials has the form $\tilde{\mu} = \begin{bmatrix} \mu_r & i\kappa & 0 \\ -i\kappa & \mu_r & 0 \\ 0 & 0 & 1 \end{bmatrix}$, where $\mu_r = 1 + \frac{(\omega_0 + i\alpha\omega)\omega_m}{(\omega_0 + i\alpha\omega)^2 - \omega^2}$, $\kappa = \frac{\omega\omega_m}{(\omega_0 + i\alpha\omega)^2 - \omega^2}$, $\omega_m = 4\pi\gamma M_s$, $\omega_0 = \gamma\mu_0 H_0$, and $\mu_0 H_0$ is the external magnetic field along the z direction which is considered as 0.22T, $\gamma$ = 2.8MHz/Oe is the gyromagnetic ratio, α= 0.01 is the damping coefficient, and ω is the operating frequency. The saturation magnetization has been set to $4\pi\gamma M_s$=1780Oe. Non-zero insertion loss is used in the simulation (PCB material) to match the practical experiment. Since the metasurface is periodic, we simulate a single unit cell with periodic boundary condition.

**Experimental measurements:** To measure the near-field, we employ a microwave vector network analyzer (VNA) and a near-field antenna acting as a source to provide excitation of electromagnetic surface waves, which is placed 0.5 mm above the samples. The transmission measurement is performed in the microwave regime using a horn antenna (operational range 8–15 GHz) to sweep the frequency. In order to minimize the Fabry-Pérot (FP) effect between slab and antenna in the direction of transmission and reflection, the measured signal in frequency domain was transformed into time domain. Then we use time gating to isolate a single response in space to alleviate the interference pattern presented in frequency domain.

# Acknowledgments

This work was supported by the New Cornerstone Science Foundation, the Research Grants Council of Hong Kong (AoE/P-502/20 and 17309021).

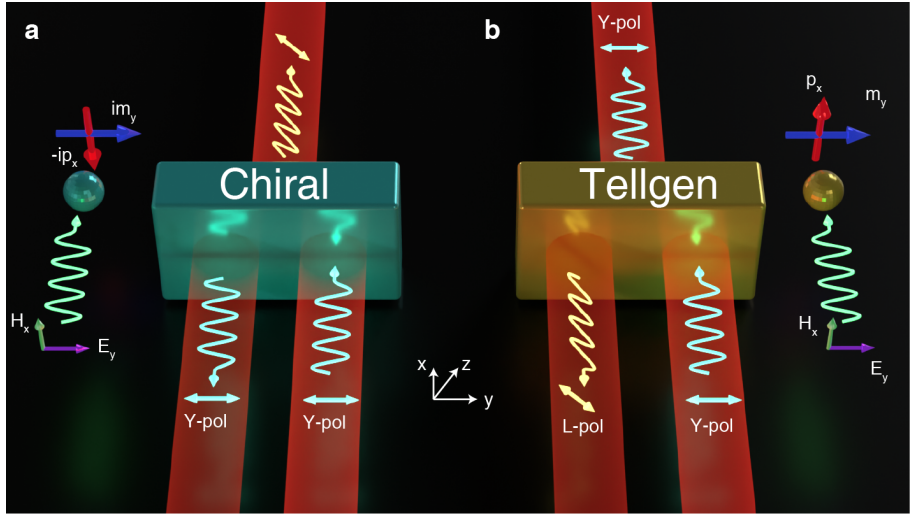

**Fig. 1. Principle of chiral and Tellegen response.** Schematics of the electromagnetic response of (**a**) a chiral and (**b**) a Tellegen metamaterial slab. The insets adjacent to each schematic display the cross-coupling polarizability of a chiral meta-atom (green sphere) and a Tellegen meta-atom (brown sphere), respectively. Here only dipole moments resulted from the cross-coupling are shown for clarity. The metamaterial slabs positioned in the *x-y* plane, are subjected to normal illumination by a plane wave propagating along the *z*-direction. The labels "Y-pol" and "L-pol" correspond to y-polarization and linear polarization, respectively.

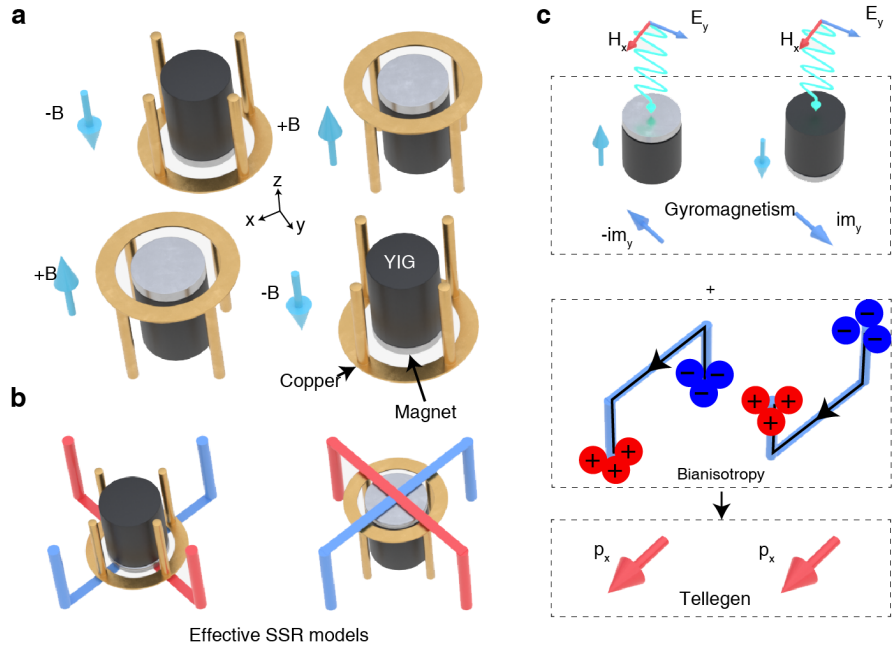

**Fig. 2 The design of the metamaterial unit-cell with the pure Tellegen response. a,** Schematic of a Tellegen unit-cell comprised four sub-particles that are constructed by inserting an YIG rod (black cylinder) in the saddle-shaped metallic coil (yellow color). Two sub-particles invert the orientation, biased by opposite external magnetic fields, as denoted by blue arrows. **b,** The saddle-shaped metallic coil of the sub-particles can be effectively regarded as two orthogonal splitting resonance rings (SRR) along *x*-direction (blue color) and *y*-direction (red color), respectively. **c,** The underlying mechanism responsible for generating the pure Tellegen response. Upon the incidence of a plane wave with *y*-polarization, the *x*-oriented magnetic field $H_x$ induces a magnetic dipole $-im_y$ in the YIG rod, and then excites the electric current along the SRR in the *x*-direction, inducing the electric dipole $p_x$. The current and charge accumulation are shown in the SRR. The inversed structure cancels the unwanted gyromagnetic and bianisotropic response.

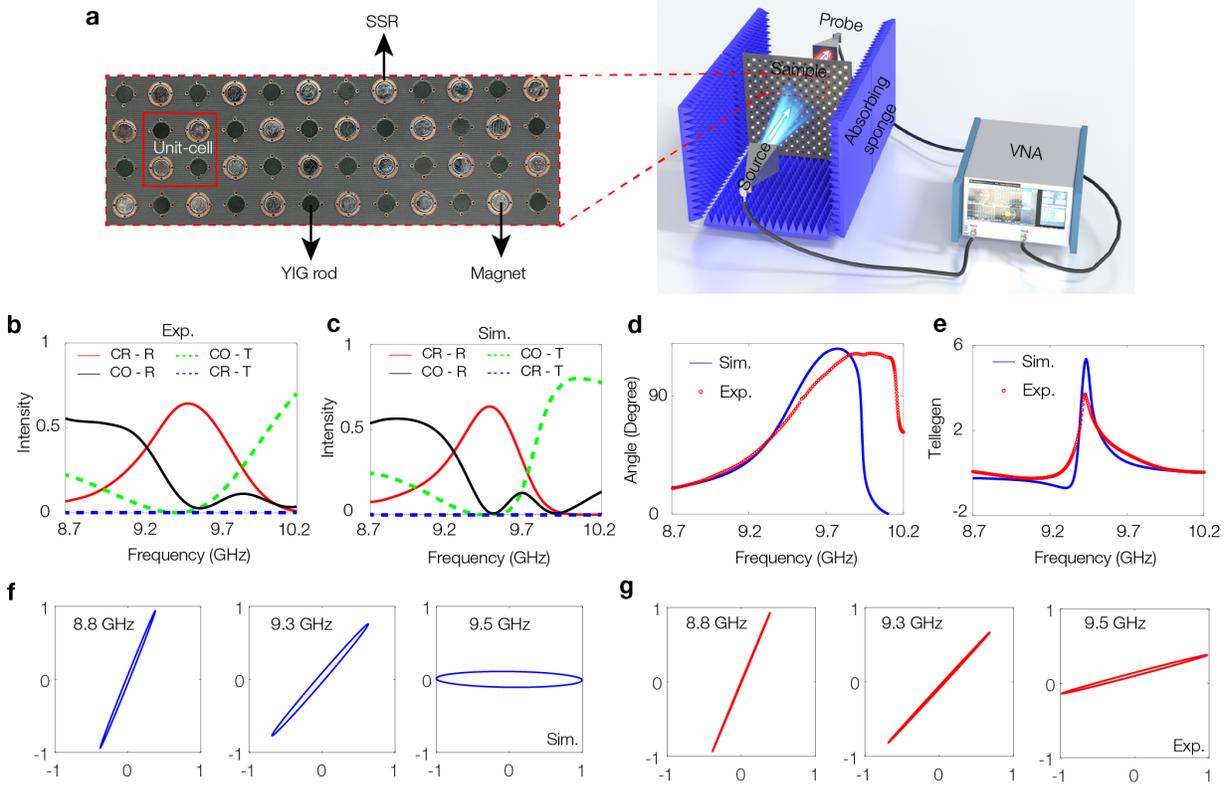

**Fig. 3. Experimental measurement of the Tellegen slab. a,** The experimental setup for measuring the transmission, utilizing antennas connected to a Vector network analyzer (VNA). The upper panel displays a zoom-out portion of the actual sample, with the silver and black cylinders representing the magnets and YIG rods inserted into the circuit board, respectively. The red frame highlights a unit-cell structure. **b-c**, Measured and simulated reflection and transmission intensity spectra. The labels "CO-" and "CR-" indicate co-polarization and cross-polarization respectively, and "T" and "R" present transmission and reflection respectively. **d**, Simulated (lines) and measured (symbols) Kerr rotation angle induced by the Tellegen response. **e,** Extracted effective Tellegen parameter obtained from simulated (line) and measured (symbols) results. **f-g,** the simulated and measured polarization state of the reflected beam at the specific frequencies.

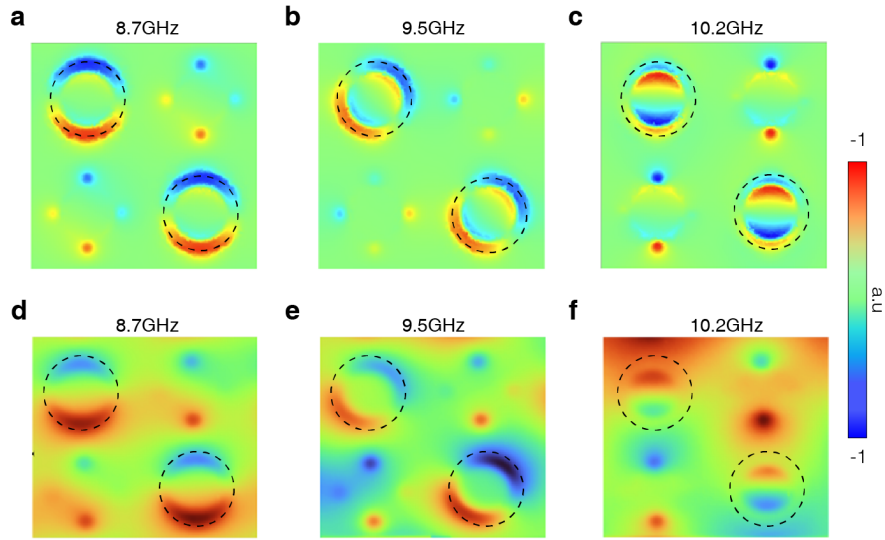

**Fig. 4. Near-field characterization of the Tellegen metamaterials slab.** The **(a-c)** simulated and **(d-f)** measured $E_z$ component of the nearfield for a metamaterial unit-cell at three different frequencies: 8.7 GHz, 9.5 GHz, and 10.2 GHz. The center of the circular metal coil is plotted by dashed line.